\documentstyle[aps,pra]{revtex}
\title{A Continuous Transition Between Quantum and
Classical Mechanics (I)} \author{Partha Ghose} \address{S. N. Bose
National Centre for Basic Sciences, JD/III, Salt Lake, Kolkata 700
098, India}
\newcommand{\be}{\begin{equation}}
\newcommand{\ee}{\end{equation}}
\newcommand{\ben}{\begin{eqnarray}}
\newcommand{\een}{\end{eqnarray}}
\newcommand{\bF}{\begin{figure}}
\newcommand{\eF}{\end{figure}}
\begin{document}
\maketitle

\begin{abstract}

In spite of its popularity, it has not been possible to vindicate the
conventional wisdom that classical mechanics is a limiting case of quantum
mechanics. The purpose of the present paper is to offer an alternative
formulation of classical mechanics which provides a continuous transition
to quantum mechanics via environment-induced decoherence.

\end{abstract}


\section{Introduction}

One of the most puzzling aspects of quantum mechanics is the quantum
measurement problem which lies at the heart of all its interpretations.
Without a measuring device that functions classically, there are no `events'
in quantum mechanics which postulates that the wave function contains
{\it complete} information of the system concerned and evolves linearly and
unitarily in accordance with the Schr\"{o}dinger equation. The system cannot
be said to `possess' physical properties like position and momentum
irrespective of the context in which such properties are measured. The
language of quantum mechanics is not that of realism.

According to Bohr the classicality of a measuring device is {\it
fundamental} and cannot be {\it derived} from quantum theory. In
other words, the process of measurement cannot be analyzed within
quantum theory itself. A similar conclusion also follows from von
Neumann's approach \cite{vN}. In both these approaches the border
line between what is to be regarded as quantum or classical is,
however, arbitrary and mobile. This makes the theory intrinsically
ill defined.

Some recent approaches have attempted to {\it derive} the
classical world from a quantum substratum by regarding quantum
systems as open. Their interaction with their `environment' can be
shown to lead to effective {\it decoherence} and the emergence of
quasi- classical behaviour \cite{Joos}, \cite{Zurek1}. However,
the very concepts of a `system' and its `environment' already
presuppose a clear cut division between them which, as we have
remarked, is mobile and ambiguous in quantum mechanics. Moreover,
the reduced density matrix of the `system' evolves to a diagonal
form only in the pointer basis and not in the other possible bases
one could have chosen. This shows that this approach does not lead
to a real solution of the measurement problem, as claimed by Zurek
\cite{Zurek2}, though it is an important development that sheds
new light on the emergence of quasi-classical behaviour from a
quantum susbstratum.

The de Broglie-Bohm approach \cite{BH}, on the other hand, does not accept the
wave function description as complete.
Completeness is achieved by introducing the position of the particle as an
additional variable (the so-called `hidden variable') with
an {\it ontological} status. The wave function at a
point is no longer just the probability amplitude that a particle will be
{\it found} there if a measurement were to be made, but the probability
amplitude that a particle {\it is} there even if no measurement is made.
It is a realistic description, and measurements are reduced to ordinary
interactions and lose their mystique. Also, the classical limit is much
better defined in this approach through the `quantum potential' than in
the conventional approach. As a result, however, a new problem is unearthed,
namely, it becomes quite clear that classical theory admits ensembles of a
more general kind than can be reached from standard quantum ensembles.
The two theories are really disparate while having a common domain of
application \cite{Holland}.

Thus, although it is tacitly assumed by most physicists that
classical physics is a limiting case of quantum theory, it is by
no means so. Most physicists would, of course, scoff at the
suggestion that the situation may really be the other way round,
namely, that quantum mechanics is contained in a certain sense in
classical theory. This seems impossible because quantum mechanics
includes totally new elements like $\hbar$ and the uncertainty
relations and the host of new results that follow from them. Yet,
a little reflection shows that if true classical behaviour of a
system were really to result from a quantum substratum through
some process analogous to `decoherence', its quantum behaviour
ought also to emerge on isolating it sufficiently well from its
environment, i.e., by a process which is the `reverse of
decoherence'. In practice, of course, it would be impossible to
reverse decoherence once it occurs for a system. Nevertheless, it
should be possible in principle to think of such a process. In
fact, it is possible to prepare a system sufficiently well
isolated from its environment so that its quantum behaviour can be
observed. If this were not possible, it would have been impossible
ever to observe the quantum features of any system.

Thus, it would appear that there must be a continuous link between
classical and quantum mechanics bridged by environment-induced
decoherence, although no one has so far succeeded in formulating
this link clearly and satisfactorily. The purpose of this paper is
to show that it is possible to formulate such a link.

We are keenly aware that we are putting forward a view that is
clearly heretic within the reigning paradigm. But it is based on
the following general considerations:

1. We accept Bell's criticism that quantum mechanics is an
inherently ambiguous theory because of the measurement problem
which nobody has been able to solve in eventy six years without
introducing nonlocal hidden variables or additional terms in the
Schr\"odinger equation (as in the GRW model) or invoking many
universes.

2. We accept Bohr's point of view that the classical world cannot
be {\it derived} from quantum mechanics but is essential for its
interpretation.

3. We see the virtue of the idea of decoherence as a possible
bridge between the classical and quantum worlds, but reject the
claim that it solves the measurement problem.

4. We assume the world is real in the classical sense, and that
both classical and (Bohmian) quantum mechanics are accurate
descriptions of nature in appropriate limits.

Our proposal must be viewed within the above framework of
assumptions.

\section{The Hamilton-Jacobi Theory}

Our starting point is the non-relativistic Hamilton-Jacobi equation

\be
\partial S_{cl}/\partial t +  \frac{(\,\nabla\,S_{cl})^2}{2 m}\,+
\,V(x) = 0
\label{eq:a}
 \ee
for the action $S_{cl}$ of a classical paticle in an external potential $V$,
together with the definition of the momentum

\be
{\bf p}  = m\,\frac{d {\bf x}}{d t}
= {\bf \nabla} S_{cl}
\label{eq:b}
\ee
and the continuity equation

\be
\frac{\partial \rho_{cl} ( {\bf x}, t )}{\partial t} +
{\bf \nabla}\,.\, (\,\rho_{cl}\, \frac{{\bf \nabla}\,S_{cl} }{m}) = 0
\label{eq:c}
\ee
for the position distribution function $\rho_{cl} ( {\bf x}, t )$ of the
ensemble of trajectories
generated by solutions of equation (\ref{eq:a}) with different initial
conditions (position or momentum).
Suppose we introduce a complex wave function

\be
\psi_{cl}\,(\,{\bf x}\,,\,t\,) = R_{cl}\,(\,{\bf x}\,,\,t\,)\,
exp\,(\,{\frac{i}{\hbar}\,S_{cl}})
\label{eq:d}
\ee
into the formalism by means of the equation

\be
\rho_{cl}\,(\,{\bf x}\,,\,t\,) = \psi_{cl}^*\,\psi_{cl} = R_{cl}^2\,.
\label{eq:e}
\ee
What is the equation that this wave function must satisfy such that the
fundamental equations (\ref{eq:a}) and (\ref{eq:c}) remain unmodified?
The answer turns out to be the modified Schr\"{o}dinger equation
\cite{Holland}

\be
i\hbar\,\frac{\partial \psi_{cl}}{\partial t} = \left(-\frac{\hbar^2}{2 m}\,
\nabla^2  + V(x)\right)\,\psi_{cl} - Q_{cl}\,\psi_{cl}
\label{eq:f}
\ee
where

\be
Q_{cl} = - \frac{\hbar^2}{2 m}\,\frac{\nabla^2 R_{cl}}{R_{cl}}
\label{eq:g}
\ee
Thus, a system can behave classically in spite of it having an associated
wave function that satisfies this modified Schr\"{o}dinger equation.

Notice that the last term in this equation is nonlinear in
$\vert\psi_{cl}\vert$, and is {\it uniquely} determined by the requirement that
all quantum mechanical effects such as superposition, entanglement and
nonlocality be eliminated. It is therefore to be sharply
distinguished from certain other types of nonlinear terms that have been
considered in constructing nonlinear versions of quantum mechanics
\cite{Weinberg}.
An unacceptable consequence of such nonlinear terms (which are, unlike
$Q_{cl}$, bilinear
in the wave function) is that superluminal signalling using quantum
entanglement becomes possible in such theories \cite{Gisin}. Since
$Q_{cl}$ eliminates quantum superposition and entanglement, it cannot
imply any such possibility. Usual action-at-a-distance is, of course,
implicit in non-relativistic mechanics, and can be eliminated in a Lorentz
invariant version of the theory, as we will see later.

Deterministic nonlinear terms with arbitrary parameters have also been
introduced in the Schr\"{o}dinger equation to bring about
collapse of quantum correlations \cite{GRW} for isolated macroscopic
systems. Such terms also imply superluminal signals via quantum entanglement.
The term
$Q_{cl}$ is different from such terms as well in that it
has no arbitrary parameters in it and eliminates
quantum correlations for all systems deterministically, irrespective
of their size.

{\it Most importantly, it is clear from the above analysis that
none of the other types of nonlinearity can guarantee strictly
classical behaviour described by equations (\ref{eq:a}) and
(\ref{eq:c}).}

Notice that the additional interaction of a classical system with
its environment in the form of the effective potential $Q_{cl}$
becomes manifest only when the Hamilton-Jacobi equation is recast
in terms of the classical wave function (equations (\ref{eq:f})
and (\ref{eq:g})). This is why the Hamilton-Jacobi equation can be
written without ever knowing about this interaction. The wave
function approach reveals what lies hidden and sterile in the
traditional classical approach. This is a significant new insight
offered by the wave function approach.

\section{Bohmian Mechanics}

The wave function $\psi$  of a quantum mechanical
system, on the other hand, must of course satisfy the
Schr\"{o}dinger equation
\be
i\,\hbar\,\frac{\partial \psi}{\partial t} = -\frac{\hbar^2}{2m}\,\nabla^2 \psi
+ V\,\psi\,\,.
\label{eq:i}
\ee
Using a polar representation similar to (\ref{eq:d}) for $\psi$ in this
equation and separating
the real and imaginary parts, one can now derive the {\it modified}
Hamilton-Jacobi equation

\be
\partial S/\partial t + \frac{(\nabla S)^2}{2m} + Q + V = 0
\label{eq:j}
\ee
for the phase $S$ of the wave function, where $Q$ is given by

\be
Q = - \frac{\hbar^2}{2 m}\,\frac{\nabla^2 R}{R}\,,
\label{eq:g2}
\ee
and the continuity equation

\be
\frac{\partial \rho ( {\bf x}, t )}{\partial t} +
{\bf \nabla}\,.\, (\,\rho\, \frac{{\bf \nabla}\,S }{m}) = 0
\label{eq:c2}
\ee
These differential equations ((\ref{eq:j}) and (\ref{eq:c2})) now become
coupled differential equations which determine $S$ and $\rho = R^2$.
Note that {\it the phase $S$ of a quantum mechanical system
satisfies a modified
Hamilton-Jacobi equation with an additional potential $Q$ called the
``quantum potential''.} Its properties are therefore different from
those of the
classical action $S_{cl}$ which satisfies equation (\ref{eq:a}) .
Applying the operator ${\bf \nabla}$ on equation
(\ref{eq:j}) and using the definition of the momentum (\ref{eq:b}), one
obtains the equation of motion

\be
 \frac{d \bf p}{d t} = m\,\frac{d^2\,{\bf x}}{d t^2} = -\, {\bf \nabla}\,(\,V + Q)
\label{eq:k}
\ee
for the quantum particle. Integrating this equation or, equivalently
equation (\ref{eq:b}), one obtains the Bohmian trajectories $x(t)$ of the
particle corresponding to different initial positions.
The departure from the classical Newtonian
equation due to the presence of the ``quantum potential'' $Q$ gives
rise to all the
quantum mechanical phenomena such as the existence of discrete stationary
states, interference phenomena, nonlocality and so on. This agreement with
quantum mechanics is achieved by requiring that the initial
distribution $P$ of the particle is given
by $R^2 (\,x(t)\,, 0\,)$. The continuity
equation (\ref{eq:c2}) then guarantees that it will agree with $R^2$ at
all future times.
This guarantees that the
averages of all dynamical
variables of the particle taken over a Gibbs ensemble of its trajectories
will always agree with
the expectation
values of the corresponding hermitian operators in standard
quantum mechanics.
This is essentially the de Broglie-Bohm quantum theory of
motion. For further details about this theory and its relationship
with standard
quantum mechanics, the reader is referred to the comprehensive book by
Holland \cite{Holland} and the one by Bohm and Hiley \cite{BH}.

\section{Environment Induced Decoherence}

Now, let us for the time being assume that quantum mechanics is the more
fundamental theory from which classical mechanics follows in some limit.
Consider a quantum mechanical system interacting with its
environment. It evolves according to the Schr\"{o}dinger equation

\be
i\,\hbar\,\frac{\partial \psi}{\partial t} = \left(-\frac{\hbar^2}{2 m}\,
\nabla^2  + V(x)  + W \,\right)\,\psi
\label{eq:h}
\ee
where $W$ is the potential due to the environment experienced
by the system. For a complex enough environment such as a heat bath,
the density matrix of the system in the position representation
quickly evolves to a diagonal form. In a special model in which a
particle interacts only with the thermal excitations of a scalar field
in the high temperature limit, the density matrix evolves
according to the {\it master equation} \cite{Zurek}

\be
\frac{d \rho}{d t} = - \gamma ( x - x^{'} )
( \partial_{x} - \partial_{x^{'}} ) \rho -
\frac{2 m \gamma k_{B} T}{\hbar^2} ( x - x^{'})^2 \rho
\ee
where $\gamma$ is the relaxation rate, $k_{B}$ is the Boltzmann
constant and $T$ the temperature of the field.
It follows from this equation that quantum coherence falls off at
large separations as the square of $\Delta x = (x - x^{'})$. The
decoherence time scale is given by

\be
\tau_{D} \approx \tau_{R} \frac{\hbar^2}{2 m k_{B} (\Delta x)^2} =
\gamma^{-1} \biggl( \frac{\lambda_{T}}{\Delta x} \biggr)^2
\label{eq:X}
\ee
where $\lambda_{T} = \hbar/\sqrt{2 m k_{B} T}$ is the thermal de Broglie
wavelength and $\tau_{R} = \gamma^{-1}$. For a macroscopic object
of mass $m = 1$ g at room temperature ( $T = 300 K$) and separation
$\Delta x = 1$ cm, the ratio $\tau_{D}/\tau_{R} = 10^{- 40}$ ! Thus,
even if the relaxation time was of the order of the age of the universe,
$\tau_{R} \simeq 10^{17}$ sec, quantum coherence would be destroyed in
$\tau_{D} \simeq 10^{- 23}$ sec. For an electron, however, $\tau_{D}$
can be much more than $\tau_{R}$ on atomic and larger scales.

However, the diagonal matrix does not become diagonal in, for
example, the momentum representation, showing that all coherence
has not altogether been destroyed. The FAPP diagonal density
matrix does not therefore correspond to the classical limit in
which it should be diagonal in both representations. In addition,
it does not correspond to a {\it heterogeneous} ensemble and the
measurement problem remains.

This is not hard to understand once one realizes that a true
classical system must be governed by a Schr\"{o}dinger equation
that is {\it modified} by the addition of a unique term that is
nonlinear in $|\psi |$ (equation (\ref{eq:f})), and that {\it such
a nonlinear term cannot arise from unitary Schr\"{o}dinger
evolution.}  On the contrary, it is not unnatural to expect a
linear equation of the Schr\"{o}dinger type to be the limiting
case of a nonlinear equation like equation (\ref{eq:f}). It is
therefore tempting to interpret the last term in equation
(\ref{eq:f}) as an `effective' potential that represents the
coupling of the classical system to its environment. It is
important to bear in mind that in such an interpretation, the
potential $Q_{cl}$ must obviously be regarded as {\it
fundamentally given} and {\it not derivable from a quantum
mechanical substratum}, being uniquely and solely determined by
the requirement of classicality, as shown above.

\section{The Classical Wavefunction}

Let us now consider a quantum system which is inserted into
a thermal bath at time $t = 0$. If it is to evolve into a genuinely classical
system after a sufficient lapse of time $\Delta t$, its wave function
$\psi$ must satisfy the equation of motion

\ben i\,\hbar\,\frac{\partial \psi}{\partial t} =
\left(-\frac{\hbar^2}{2 m}\, \nabla^2  + V(x)  - \lambda (t)
Q_{cl} \,\right)\,\psi \label{eq:h1} \een where $\lambda (0)
\rightarrow 0$ is the quantum limit and $\lambda (\Delta t) = 1$
is the classical limit. (Here $\Delta t \gg\tau_{D}$ where
$\tau_{D}$ is typically given by $\gamma^{-1} (\lambda_{T}/\Delta
x )^2$ (\ref{eq:X}).) Thus, for example, if $ =\lambda (t) = 1 -
exp ( -t /\tau_{D} )$, a macroscopic system would very rapidly
behave like a true classical system at sufficiently high
temperatures, whereas a mesoscopic system would behave neither
fully like a classical system nor fully like a quantum mechanical
system at appropriate temperatures for a much longer time. What
happens is that the reduced density operator of the system evolves
according to the equation

\ben
\rho (x, x^{'}, \Delta t ) &=& exp (- i \int_{0}^{\Delta t}\lambda Q_{cl}
d t/\hbar)
\rho (x, x^{'},  0)
exp ( i \int_{0}^{\Delta t}\lambda Q_{cl} d t/\hbar )\\
&=& R^2 ( x, \Delta t ) \delta^{3} ( x - x^{'} )
\een
during the time interval $\Delta t$ during which the nonlinear interaction
$\lambda Q_{cl}$ completely destroys
all superpositions, so that at the end of this time interval the system
is fully classical and the equation
for the density operator reduces to the Pauli master equation for a
classical system.

A variety of functions $\lambda (t)$ would satisfy the requirement
$\lambda=0$ and $\lambda=1$. This is not surprising and is
probably a reflection of the diverse ways in which different
systems decohere in different environments. We will elaborate on
this in the following paper.

It is clear that a system must be extremely well isolated
($\lambda \rightarrow 0$) for it to behave quantum mechanically.
Such a system, however, would inherit only a de Broglie-Bohm
ontological and causal interpretation, not an interpretation of
the Copenhagen type. The practical difficulty is that once a
quantum system and its environment get coupled, it becomes FAPP
impossible to decouple them in finite time because of the
extremely large number of degrees of freedom of the environment.
However, we know from experience that it is possible to {\it
create} quantum states in the laboratory that are very well
isolated from their environment. Microscopic quantum systems are,
of course, routinely created in the laboratory (such as single
atoms, single electrons, single photons, etc.,) and considerable
effort is being made to create isolated macroscopic systems that
would show quantum coherence, and there is already some evidence
of the existence of mesoscopic `cat states' which decohere when
appropriate radiation is introduced into the cavity \cite{Brune}.

Equation (\ref{eq:h1}) is a new equation that is different from
both the Schr\"odinger equation and the wave equation that
describes classical mechanics (\ref{eq:f}), and provides a smooth
link between the two. We supplement this equation with the
postulate that $\psi$ is single-valued. Then it would contain
quantum mechanics in the limit $\lambda \rightarrow 0$. It would
also contain classical mechanics in the limit $\lambda =1$,
although the single-valuedness of the wavefunction is not a
necessary requirement for this limit. It can therefore form a
sound starting point for studying  mesoscopic systems in a new way
in which they are parametrized by $0 < \lambda \leq 1$ and lie
anywhere in the continuous spectrum stretching between the quantum
and classical limits.  We will show in the following
paper\cite{meso} that this leads to new physical predictions for
mesoscopic systems that cannot be obtained from either standard
quantum mechanics or standard Bohmian mechanics which accepts the
Schr\"odinger equation as fundamental and not our new equation
(\ref{eq:h1}).

\section{The Klein-Gordon Equation}

Let the Hamilton-Jacobi equation for free relativistic classical particles
be

\be
\frac{\partial S_{cl}}{\partial t} + \sqrt{ (\partial_i S_{cl})^2\,c^2 + m_0^2\,c^4}
= 0\,.
\label{eq:n}
\ee
Then, using the relation $p_\mu = - \partial_\mu S_{cl} = m_0\,u_\mu$ where
$u_\mu = d\,x_\mu/d\,\tau$ with $\tau = \gamma^{-1}\,t$, $\gamma^{- 1}
= \sqrt{1 - v^2/c^2}, v_i = d\,x_i/d\,t$, the particle equation of motion is
postulated to be

\be
m_0\,\frac{d u_\mu}{d \tau} = 0 = \frac{d\,p_\mu}{d\,\tau} \,.
\label{eq:o}
\ee
It is quite easy to show that the classical
equations (\ref{eq:n}) and (\ref{eq:c}) continue to hold if one describes the
system in terms of a complex wave function
$\psi_{cl} = R_{cl}\, exp\,(\,\frac{i}{\hbar} S_{cl}\,)$ that satisfies the modified
Klein-Gordon equation

\be
\left(\,\Box + \frac{m_0^2\, c^2}{\hbar^2} - \frac{ Q_{cl}}{\hbar^2} \right)
\,\psi_{cl} = 0
\label{eq:p}
\ee
with

\be
Q_{cl} = \hbar^2 \frac{\Box R_{cl}}{R_{cl}}\,.
\label{eq:q}
\ee
As in the non-relativistic case, $ Q_{cl}$ may be interpreted as an effective
potential in which the system
finds itself when described in terms of the wave function $\psi_{cl}$.
If this potential goes to zero in some limit, one obtains the free Klein-Gordon
equation which is the quantum limit.

On the other hand, using $\psi = R\, exp\,(\,\frac{i}{\hbar} S\,)$ in the
Klein-Gordon equation
and separating the real and imaginary parts, one obtainds respectively
the equation

\be
\frac{1}{c^2}\,\left(\frac{\partial S}{\partial t}\right)^2
- \left( \partial_i S\right)^2 - m_0^2\, c^2 - Q = 0
\label{eq:r1}
\ee
which is equivalent to the modified Hamilton-Jacobi equation
\be
\left(\frac{\partial S}{\partial t}\right)
+ \sqrt{ \left( \partial_i S\right)^2\,c^2 + m_0^2\, c^4 + c^2\,Q}
= 0
\label{eq:r}
\ee
and the continuity equation

\be
\partial^\mu\,(\,R^2 \partial_\mu\,S\,) = 0\,.
\ee
One can then identify the four-current as $j_\mu = - R^2 \partial_\mu S$ so
that $\rho = j_0 =  R^2 E/c$ which is not positive definite because $E$ can be
either positive or negative, and therefore, as is well known,
it is not possible to interpret it as a probability density.

Nevertheless, let us note in passing that, if use is made of the definition
$p_\mu = - \partial_\mu\,S$ of the particle four-momentum, (\ref{eq:r1})
implies

\be
p_\mu\,p^\mu = m_0\,c^2 + Q
\label{eq:t}
\ee
and $p_\mu = M_0\,u_\mu$ where $M_0 = m_0\,\sqrt{1 +
Q/m_0^2\,c^2}$. Thus, the quantum potential $Q$ acts on the particles and
contributes to their energy-momentum so that they are off their mass-shell.
\footnote{The author is grateful to E. C. G. Sudarshan for drawing his
attention to this important point.}
Applying the operator $\partial_\mu$ on equation (\ref{eq:r1}), we get the
equation of motion

\be
\frac{d\,p_\mu}{d\,\tau} = \frac{\partial_\mu\,Q}{2\,M_0}
\ee
which has the correct non-relativistic limit. The equation for
the acceleration of the particle is therefore given by \cite{Holland}

\be
\frac{d\,u_\mu}{d\,\tau} = \frac{1}{2\,m_0^2}\,
(\,c^2\,g_{\mu\nu} - u_\mu\,u_\nu\,)\,\partial^\nu\,log\,(1 + \frac{Q}{m_0^2\,c^2}\,)\,.
\ee
If, on the other hand, one uses the modified Klein-Gordan equation
(\ref{eq:p}) and the corresponding Hamilton-Jacobi equation (\ref{eq:n}),
the particles are on their mass-shell and the free particle classical
equation (\ref{eq:o}) is satisfied.

\section{Relativistic spin 1/2 particles}

Let us now examine the Dirac equation for relativistic spin $1/2$ particles,

\be
(\,i \hbar \gamma_\mu \partial^\mu + m_0\, c )\,\psi = 0.
\label{eq:D1}
\ee
Let us write the components of the wave function
$\psi$ as $\psi^a = R\,\theta^a\,exp\,\,(\,\frac{i}{\hbar}\,S^a)$,
$\theta^a$ being a spinor component. It is not straightforward
here to separate the real and imaginary parts as in the previous cases.
One must therefore follow a different method for relativistic fermions.

It is well known that every component $\psi^a$ of the Dirac wave function
satisfies the Klein-Gordan equation. It follows therefore, by putting
$\psi^a = R \theta^a\,exp\,(\,i\,S^a/\hbar\,)$, that $S^a$ must satisfy the modified
Hamilton-Jacobi equation

\be
\partial_\mu\,S^a\,\partial^\mu\,S^a - m_0^2\, c^2 - Q^a = 0\,.
\label{eq:D2}
\ee
where $Q^a = \hbar^2\,\Box\,R\,\theta^a/R\,\theta^a$.
Summing over $a$, we get

\be
\sum_a\,\partial_\mu\,S^a\,\partial^\mu\,S^a - 4\,m_0^2\, c^2 -
\sum_a\,Q^a = 0\,.
\label{eq:D3}
\ee
Defining

\ben
\partial_\mu\,S\,\partial^\mu\,S &=& \frac{1}{4}\,
\sum_a\,\partial_\mu\,S^a\,\partial^\mu\,S^a\\
Q &=& \frac{1}{4}\,\sum_a\,Q^a\,,
\label{eq:D4}
\een
we have

\be
\partial_\mu\,S\,\partial^\mu\,S - m_0^2\, c^2 -
Q = 0\,.
\label{eq:D5}
\ee
Then, defining the particle four-momentum by $p_\mu = - \partial_\mu\,S$,
one has $p_\mu\,p^\mu = m_0^2\,c^2 + Q$. Therefore, one has the equation of
motion

\be
\frac{d\,p_\mu}{d\,\tau} =
\frac{\partial_\mu\,Q}{2\, M_0}\,.
\label{eq:D6}
\ee
The Bohmian 3-velocity of these particles is defined by the relation

\be
v_i = \gamma^{- 1}\,u_i = c\,\frac{u_i}{u_0} = c\,\frac{j_i}{j_0}
= c\,\frac{\psi^{\dagger}\,\alpha_i\,\psi}{\psi^{\dagger}\,\psi}\,.
\label{eq:D7}
\ee
Then, it follows that

\be
u_\mu = \gamma\,v_\mu = \gamma\,c\,\frac{j_\mu}{\rho}
\label{eq:D8}
\ee
where $\rho = \psi^{\dagger}\,\psi$. This relation is satisfied because
$j_\mu\,j^\mu = \rho^2/\gamma^2$ if (\ref{eq:D7}) holds.

As we have seen, for a classical theory of spinless particles,
the correct equation for the associated wave function is
the modified Klein-Gordon equation (\ref{eq:p}). Let the corresponding
modified wave equation for classical spin $1/2$ particles be of the form

\be
\left(\,i\, \hbar\, \gamma_\mu\, D^\mu + m_0\, c \,\right)\,\psi_{cl} = 0
\label{eq:D9}
\ee
where $D^\mu = \partial^\mu + (i/\hbar)\,Q^\mu$. Then we have

\be
(\,D_\mu\,D^\mu + \frac{m_0^2\,c^2}{\hbar^2}\,)\,\psi_{cl}^a = 0\,.
\label{eq:D10}
\ee
Writing $\psi_{cl}^a = R_{cl}\,\theta^a\,exp\,(\frac{i}{\hbar}\,S_{cl}^a )$, one obtains

\be
\partial_\mu\,S_{cl}^a\,\partial^\mu\,S_{cl}^a - m_0^2\, c^2 - Q_{cl}^a + Q_\mu\,Q^\mu
- 2\,Q_\mu\,\partial^\mu\,S_{cl}^a = 0
\label{eq:D11}
\ee
where

\be
Q^{a}_{cl}= \frac{\hbar^2 \Box R_{cl} \theta^{a}}{R_{cl} \theta^{a}}.
\ee
Define a diagonal matrix
$B_\mu^{a\,b} \equiv \partial_\mu\,S_{cl}^a\,\delta^{a\,b}$ such that

\be
\frac{1}{2}\,Tr B_\mu = \frac{1}{2}\,\sum_a\,\partial_\mu\,S_{cl}^a
\equiv \partial_\mu\,S_{cl}\,.
\ee
Then

\ben
\partial_\mu\,S_{cl}\,\partial^\mu\,S_{cl} &=& \frac{1}{4}\,Tr\,B_\mu\,Tr\,B^\mu =
\frac{1}{4}\,Tr\,(\,B_\mu\,B^\mu\,)\\&=& \frac{1}{4}\,\sum_a
\partial_\mu S_{cl}^a\,\partial^\mu S_{cl}^a\,.
\een
Therefore, taking equation (\ref{eq:D11}) and summing over $a$, we have

\be
\partial_\mu\,S_{cl}\,\partial^\mu\,S_{cl} - m_0^2\, c^2 -
Q_{cl} + Q_\mu\,Q^\mu - Q_\mu\,\partial^\mu\,S_{cl} = 0
\label{eq:D12}
\ee
where

\be
Q_{cl} = \frac{1}{4} \sum_a Q^a_{cl}\,.
\ee
In order that the classical free particle equation is satisfied, the effects
of the quantum potential must be cancelled by this additional interaction,
and one must have

\be
Q_\mu\,(\,Q^\mu - \partial^\mu\,S_{cl}\,) = Q_{cl}\,.
\ee
A solution is given by

\ben
p_\mu &=& - \partial_\mu\,S_{cl} = m_0\,u_\mu\,,\\
Q_\mu &=& \alpha\,m_0\,u_\mu
\een
with

\be
\alpha = \frac{1}{2} \pm \frac{1}{2}\,\sqrt{1 + 4\,Q_{cl}/m_0^2\,c^2}\,.
\ee

\section{Relativistic spin 0 and spin 1 particles}

It has been shown \cite{Ghose} that a consistent relativistic quantum
mechanics of spin 0 and spin 1 bosons can be developed using the
Kemmer equation \cite{Kemmer}

\be
(\,i\,\hbar\,\beta_\mu\,\partial^\mu + m_0\,c\,)\,\psi = 0\,
\label{eq:1}
\ee
where the matrices $\beta$ satisfy the algebra

\be
\beta_{\mu}\,\beta_{\nu}\,\beta_{\lambda} + \beta_{\lambda}\,\beta_{\nu}\,
\beta{\mu} = \beta_{\mu}\,g_{\nu \lambda} + \beta_{\lambda}\,g_{\nu \mu}\,.
\label{eq:2}
\ee
The $5\times 5$ dimensional representation of these matrices describes spin 0
bosons and the $10 \times 10$ dimensional representation describes spin 1
bosons. Multiplying (\ref{eq:1}) by $\beta_0$, one obtains the
Schr\"{o}dinger form of the equation

\be
i\,\hbar\,\frac{\partial \psi}{d t} = [\,- i\,\hbar\,c\, \tilde{\beta}_i\,
\partial_i - m_0\,c^2\,\beta_0\,]\,\psi
\label{eq:3}
\ee
where $\tilde{\beta}_i \equiv \beta_0\,\beta_i - \beta_i\,\beta_0$. Multiplying
(\ref{eq:1}) by $1- \beta_0^2$, one obtains the first class constraint

\be
i\,\hbar\,\beta_i\,\beta_0^2\,\partial_i\,\psi = -m_0\,c\,(\,1 - \beta_0^2\,)
\,\psi.
\label{eq:4}
\ee
The reader is referred to Ref. \cite{Ghose} for further discussions regarding
the significance of this constraint.

If one multiplies equation (\ref{eq:3}) by $\psi^{\dagger}$ from the left,
its hermitian conjugate by $\psi$ from the right and adds the resultant
equations, one obtains the continuity equation

\be
\frac{\partial\,( \psi^{\dagger}\,\psi )}{\partial t} + \partial_i\,
\psi^{\dagger}\,\tilde{\beta}_i\,\psi = 0\,.
\ee
This can be written in the form

\be
\partial^\mu\,\Theta_{\mu 0} = 0
\ee
where $\Theta_{\mu \nu}$ is the symmetric energy-momentum tensor with
$\Theta_{0 0} = - m_0\,c^2\,\psi^{\dagger}\,\psi < 0$. Thus, one can define
a wavefunction $\phi = \sqrt{m_0\,c^2/E}\,\psi$ (with $E =- \int\,\Theta_{0 0}
\,dV$ ) such that
$\phi^{\dagger}\,\phi$ is non-negative and normalized and can be
interpreted as a probability density. The conserved probability current
density is $s_\mu = - \Theta_{\mu 0}/E = (\,\phi^{\dagger}\,\phi,
- \phi^{\dagger}\,\tilde{\beta}_i\,\phi )$ \cite{Ghose}.

Notice that according to the equation of motion (\ref{eq:3}), the velocity
operator for massive bosons is $c\,\tilde{\beta}_i$, so that the Bohmian
3-velocity can be defined by

\be
v_i = \gamma^{- 1}\,u_i = c\,\frac{u_i}{u_0} = c\,\frac{s_i}{s_0}
= c\,\frac{\psi^{\dagger}\,\tilde{\beta}_i\,\psi}{\psi^{\dagger}\,\psi}\,.
\label{eq:5}
\ee

Exactly the same procedure can be followed for massive bosons as for massive
fermions to determine the quantum potential and the Bohmian trajectories, except
that the sum over $a$ has to be carried out only over the independent degrees
of freedom (six for $\psi$ and six for $\bar{\psi}$ for spin-1 bosons).
The constraint (\ref{eq:4}) implies the four conditions $\vec{A} =
\vec{\nabla}\times\,\vec{B}$ and  $\vec{\nabla}\,.\,\vec{E} = 0$.

The theory of massless spin 0 and spin 1 bosons cannot be obtained simply by
taking the limit $m_0$ going to zero. One has to start with the equation
\cite{HC}

\be
i\,\hbar\,\beta_\mu \partial^\mu\,\psi + m_0\,c\,\Gamma\,\psi = 0
\label{eq:8}
\ee
where $\Gamma$ is a matrix that satisfies the following conditions:

\ben
\Gamma^2 &=& \Gamma\,\\
\Gamma\,\beta_\mu + \beta_\mu\,\Gamma &=& \beta_\mu\,.
\label{eq:9}
\een
Multiplying (\ref{eq:8}) from the left by $1 - \Gamma$, one obtains

\be
\beta_\mu\,\partial^\mu\, (\,\Gamma\,\psi\,) = 0\,.
\label{eq:10}
\ee
Multiplying (\ref{eq:8}) from the left by $\partial_{\lambda}\,
\beta^{\lambda}\,\beta^{\nu}$, one also obtains

\be
\partial^{\lambda}\,\beta_{\lambda}\,\beta_\nu\,(\,\Gamma\,\psi\,) =
\partial_\nu\, (\,\Gamma\,\psi\,)\,.
\label{eq:11}
\ee
It follows from (\ref{eq:10}) and (\ref{eq:11}) that

\be
\Box\,\, (\,\Gamma\,\psi\,) = 0
\label{eq:12}
\ee
which shows that $\Gamma\,\psi$ describes massless bosons. The Schr\"{o}dinger
form of the equation

\be
i\,\hbar\,\frac{\partial\, (\,\Gamma\,\psi\,)}{d t} = - i\,\hbar\,c
\tilde{\beta}_i\,\partial_i\, (\Gamma\,\psi)
\label{eq:13}
\ee
and the associated first class constraint

\be
i\,\hbar\,\beta_i\,\beta_0^2\,\,\partial_i\,\psi
+ m_0\,c\,(\,1 - \beta_0^2\,)\,\Gamma\,\psi = 0
\label{eq:14}
\ee
follow by multiplying (\ref{eq:8}) by $\beta_0$ and $1 - \beta_0^2$
respectively. The rest of the arguments are analogous to the massive case.
For example, the Bohmian 3-velocity $v_i$ for massless bosons can be defined by
equation (\ref{eq:5}).

Neutral massless spin-1 bosons have a special significance in physics. Their
wavefunction is real, and so their charge current $j_\mu =
\phi^{T}\,\beta_\mu\,\phi$ vanishes. However,
their probability current density $s_\mu$ does not vanish.
Furthermore, $s_i$ turns out to be proportional to the Poynting vector, as it
should.

Modifications to these equations can be introduced as
in the massive case to obtain a classical theory of massless bosons.

\section{The Gravitational Field}

Exactly the same procedure can also be applied to the gravitational field described
by Einstein's equations

\be
R_{\mu\nu} - \frac{1}{2}\,g_{\mu\nu}\,R = 0
\label{eq:a'}
\ee
for the vacuum, where $R_{\mu\nu}$ is the Ricci tensor and $R$ the curvature
scalar. In this section, following \cite{WD}, we will use the
signature $- + + +$ and the absolute system of units $\hbar = c = 16\, \pi\, G
= 1$. The decompostion of the metric is given by \cite{Holland}

\ben
ds^2 &=& g_{\mu\nu}\,d x^\mu\,dx^\nu\nonumber\\
&=& (\,N_i\,N^i - N^2\, )\,d t^2 + 2\,N_i\,dx^i\,d t + g_{ij}\,d x^i\,dx^j
\een
with $g_{i\,j}({\bf x})$, the 3-metric of a 3-surface embedded in space-time,
evolving dynamically in superspace, the space of all 3-geometries.

By quantizing the Hamiltonian constraint, one obtains in the standard
fashion the Wheeler-DeWitt equation \cite{WD}

\be
\left[\,G_{i\, j\, k\, l}\,\frac{\delta^2}{\delta g_{i\,j}\,\delta g_{k\,l}}
+ \sqrt{g}\,\,\,^3R\,\right]\,\Psi = 0
\label{eq:b'}
\ee
where $g =$ det $g_{i\,j}$, $^3R$ is the intrinsic curvature, $G_{i\,j\,k\,l}$
is the supermetric, and $\Psi [g_{i\,j}(x)]$ is a wave functional in
superspace. Substituting $\Psi = A\,\, exp\,\,(i\, S)$, one obtains as usual a
conservation law

\be
G_{i\,j\,k\,l}\,\frac{\delta}{\delta g_{i\,j}}\,\left(\,A^2\,\frac{\delta S}
{\delta g_{k\,l}}\,\right) = 0
\ee
and a modified Einstein-Hamilton-Jacobi equation

\be
G_{i\,j\,k\,l}\,\frac{\delta S}{\delta g_{i\,j}}\,\frac{\delta S}
{\delta g_{k\,l}} - \sqrt{g}\,\,\,^3R + Q = 0
\ee
where

\be
Q =  - A^{- 1}\,G_{i\,j\,k\,l}\,\delta^2\,A/\delta g_{i\,j}\,\delta g_{k\,l}
\label{eq:Z}
\ee
is the quantum potential. It is invariant under 3-space diffeomorphisms.
The causal interpretation of this {\it field} theory (as distinct from particle
mechanics considered earlier) assumes that the universe whose quantum state
is governed by equation (\ref{eq:b'}) has a definite 3-geometry at each
instant, described by the 3-metric
$g_{ij}({\bf x}, t)$ which evolves according to the classical Hamilton-Jacobi
equation

\be
\frac{\partial g_{i\,j}({\bf x}, t)}{\partial t} = \partial_i N_j
+ \partial_j N_i + 2\,N\,G_{i\,j\,k\,l}\,\frac{\delta S}
{\delta g_{k\,l}}\vert_{g_{i\,j}(x) = g_{i\,j}({\bf x},t)}
\label{eq:c'}
\ee
but with the action $S$ as a phase of the quantum wave functional.
This equation can be solved if the initial data $g_{i\,j}({\bf x}, 0)$
are specified. The metric in this field theory clearly corresponds to the
position in particle mechanics, equation (\ref{eq:c'}) being its guidance
condition.

It is now clear that one can modify the Wheeler-DeWitt equation (\ref{eq:b'})
to the form

\be
\left[\,G_{i\, j\, k\, l}\,\frac{\delta^2}{\delta g_{i\,j}\,\delta g_{k\,l}}
+ \sqrt{g}\,\,\,^3R - Q_{cl}\,\right]\,\Psi_{cl} = 0
\ee
where $Q_{cl}$ is defined by an expression analogous to (\ref{eq:Z}) with $A$
and $S$ replaced by the classical variables $A_{cl}$ and $S_{cl}$. This leads
to the classical Einstein-Hamilton-Jacobi equation

\be
G_{i\,j\,k\,l}\,\frac{\delta S_{cl}}{\delta g_{i\,j}}\,\frac{\delta S_{cl}}
{\delta g_{k\,l}} - \sqrt{g}\,\,\,^3R = 0\,.
\ee
The term $Q_{cl}$ can then be interpreted, as before, as a potential arising due to
the coupling of gravitation with other forms of energy. If this coupling could
be switched off, quantum gravity effects would become important.
The question arises as to whether this can at all be done for gravitation.

\section{Multi-particle Generalization}

The proper multi-particle generalization of eq.(\ref{eq:h1}) is
straightforward and given by \be i\,\hbar\,\frac{\partial
\Psi}{\partial t} = \left[- \sum_{i}^n \left(\frac{\hbar^2}{2
m_i}\, \nabla_i^2 + V(x_i, ...,x_n)\right) - \sum_{i} \lambda_i
(t) Q_i \,\right]\,\Psi \label{mpg} \ee where $\Psi$ stands for
$\psi(x_1, ..., x_n)$ and $i$ refers to the $i$th particle. These
particles can, in general, have different degrees of coupling to
their environment so that $\lambda_i \neq \lambda_j$. It follows
that those particles (or degrees of freedom) that are very
strongly coupled to their environment and behave classically have
$\lambda_i = 1$ and those that are less strongly coupled have $0
<\lambda_j < 1$. The effective $\lambda(t)$ is therefore given by
\be \lambda(t) = \frac{\sum_{i} \lambda_i (t) Q_i}{Q},\ee where
$Q=\sum^n_i Q_i$ \cite{holland2}. This determines the value of
$\lambda(t)$ in general for any multi-particle system.

Let us first consider building a reasonable physical theory of
observed systems in the universe (and not the universe itself
which we will take up later). The important point to note about
the many-particle generalization (eqn. \ref{mpg}) is that it
describes a multi-particle system $S$ interacting with its
environment $E$ (which is assumed to have infinite degrees of
freedom) {\it in a specific nonlinear manner} such that it is
obtained when $E$ is traced over. However, if one subdivides $S$
into two parts, say $S_a$ and $E_a$ which interact nonlinearly to
yield this equation for $S_a$ when $E_a$ is traced over, then $S$
can no longer be regarded as a quantum mechanical system.

Let us now consider the universe, i.e., the total system $S + E$
where $S$ is now the observable universe and $E$ the rest of the
universe. Then, since $S$ and $E$ are assumed to interact
nonlinearly to yield eqn. (\ref{mpg}) for $S$ when $E$ is traced
over, it is no longer legitimate to regard the total system $S +
E$ to be quantum mechanical, although it would appear from this
equation that $S + E$ obeys the Schr\"odinger equation if $\lambda
\equiv 0$. In order to avoid any internal inconsistency of this
type, we must omit the limit point $\lambda = 0$, i.e., quantum
mechanics must be viewed only as a limiting case of eqn.
(\ref{mpg}) and not as a fundamental theory. We take the point of
view here that there are no physical systems, including the
observable universe, that are truly closed and isolated, although
it is possible in principle to isolate a system as closely as one
desires. This assumption lies at the foundation of statistical
mechanics \cite{toda}. The total system $S + E$ is, of course,
closed by definition, but is a metaphysical concept to which,
admittedly, eqn. (\ref{mpg}) cannot be applied consistently.

\section{Concluding Remarks}

It is usually assumed that a classical system is in some sense a
limiting case of a more fundamental quantum substratum, but no
general demonstration for ensembles of systems has yet been given.
That classical and quantum systems may, on the other hand, be
linked through a more general system in which their typical
features are not fully expressed is, however, clear from the above
discussions. The limits therefore naturally share the ontology of
the link system. The nonlocal quantum potential that is
responsible for self-organization and the creation of varied
stable and metastable quantum structures, becomes fully active
only when the coupling of the part to the whole is completely
switched off. This is a clearly defined physical process that
links the classical and quantum domains.

According to this view, therefore, every quantum system is a
closed system and every classical system is an open system. The
first Newtonian law of motion therefore acquires a new
interpretation---the law of inertia holds for a system not when it
is isolated from everything else but when it interacts with its
environment to an extent that all its quantum aspects are
quenched. Various attempts to show that the classical limit of
quantum systems is obtained in certain limits, like large quantum
numbers and/or large numbers of constituents, have so far failed
\cite{Mermin}. The reason is clear---a linear equation like the
Schr\"{o}dinger equation can never describe a classical system
which is described by a {\it modified} Schr\"{o}dinger equation
with a nonlinear term. This nonlinear term must be generated
through some mechanism like the coupling of the system to its
environment. There are, of course, other purely {\it formal}
limits too (like $\hbar$ going to zero, for example) in which a
closed quantum system reduces to a classical system, as widely
discussed in the literature.

It is clear from the usual `decoherence' approach that the interaction
of a quantum system with its environment in the form of some kind of
heat bath is necessary to obtain a quasi-classical limit of quantum
mechanics. This is usually considered to be a major advance in recent
years. Such decoherence effects have already been measured in cavity
QED experiments. Decoherence effects are very important to take into
account in other critical experiments too, like the use of SQUIDs to
demonstrate the existence of Schr\"{o}dinger cat states. The failure to
observe cat states so far in such experiments shows how real these effects
are and how difficult it is to eliminate them even for mesoscopic systems.
I have taken these advances in our knowledge seriously in a
phenomenological sense and tried to incorporate them into a conceptually
consistent scheme.

The usual decoherence approach however suffers from the following
difficulty: it neither solves the measurement problem nor does it
lead to a truly classical phase space. This does not happen in the
approach advocated in this paper because of the new equation
(\ref{eq:h1}) which guarantees the emergence of a {\it
heterogeneous} ensemble and classical phase space.

A clear empirical difference must therefore exist between the
predictions of the usual decoherence approach and the approach
advocated in this paper, and this will be considered in more
detail in the following paper.

We must emphasize that equation (\ref{eq:h1}) cannot be {\it
derived} from quantum mechanics whereas the converse is true. The
simple reason is that the third term in (\ref{eq:h1}) is
nonlinear, and no such term can be generated within quantum
mechanics itself which is strictly linear.

This eqn. (\ref{eq:h1}) is consistent with all known phenomena in
the classical and quantum limits and opens up new possibilities in
the unchartered meso domain. The theory is conceptually clear and
precise and internally consistent within the framework of the
assumptions $1$ through $4$ stated in section I. It is capable of
making predictions that are in principle different from those of
conventional theory, as we will show in the next paper. It is
therefore not metaphysical.

One could speculate that eqn. (\ref{eq:h1}) is a pointer to a new
fundamental theory because it cannot be derived from either
quantum mechanics or classical mechanics. It is here, I guess,
that there would be serious differences of opinion.

\section{Acknowledgements}

The basic idea of this paper occurred to me during a discussion
with C. S. Unnikrishnan in the context of an experiment carried
out by R. K. Varma and his colleagues and Varma's theoretical
explanation of it in terms of the Liouville equation written as a
series of Schr\"{o}dinger-like equations. Unnikrishnan should not,
however, be held responsible for any defects of my particular
formulation of the idea.

The first version of the non-relativistic part of this paper was presented
at the ``Werner Heisenberg
Colloquium: The Quest for Unity: Perspectives in Physics and Philosophy''
organized by the Max Muller Bhavan, New Delhi, the National
Institute of Science, Technology and \& Development Studies, CSIR, Government
of India and the Indian Institute of Advanced Study, Shimla from 4th to
7th August, 1997 at Shimla.

The author is grateful to the Department of Science and Technology,
Government of India, for a research grant that enabled this work to be completed.

\end{document}